# The Optimal Division of the Energy Market into Zones: Comparison of Two Methodologies under Variable Wind Conditions

Karol Wawrzyniak, Michał Kłos, Grzegorz Oryńczak, Marcin Jakubek
National Centre for Nuclear Research, Świerk Computing Centre
Otwock-Świerk, Poland
K.Wawrzyniak@fuw.edu.pl

**Summary**

The energy market of Europe is under an intensive process of transformation. The main drivers for change are integration of national markets and growing use of renewable generation. Currently, the most popular market structures are uniform, nodal, and zonal pricing.

In uniform pricing, there is a single price of energy set on a national market for each hour of a day. This market structure is still used in many countries mainly due to historical reasons. In spite of its apparent simplicity, such an approach has serious disadvantages mainly due to the existence of congestions [1], since equilibrium set on the market does not take into account safety requirements of the grid. Introducing other forms of market helps to eliminate the uniform market limitations.

In nodal pricing, each location of the grid has its own price of energy, representing the locational value of energy, i.e. a cost of supplying extra 1 MW of energy to this location (node). This market structure solves congestion, since the psychical structure of the grid and its transmission limits are taken into account when the market equilibrium is searched for. However, such market arrangement is complicated and insufficiently transparent for market participants.

Zonal market, which can be thought of as a compromise of simplicity of uniform structure and accuracy of nodal one, introduces differentiation of prices between regions of different costs of supplying energy, but it maintains the transparency which the nodal market lacks. Still, there is no consensus in the literature with respect to methodology of identification of zones' number and their borders. In the proposed research we compare two competing methodologies, (i) consensus clustering of Locational Marginal Prices and (ii) congestion contribution identification [2], under the criterion of social welfare maximization, and taking into account variable wind generation. Below we discuss the two methodologies and the welfare criterion. In our previous work [1] we presented in more detail the methodology and results already obtained that are related to clustering of LMP. The research related to second methodology is in ongoing phase. Hence, the methodology is presented here but the complete results will be provided shortly.

**(i)   Consensus clustering of Locational Marginal Prices**

The first methodology is based on *nodal prices,* called also Locational Marginal Prices (LMP) [3],[4]. Nodal pricing is a method of determining prices in all locations of the transmission grid. The price at each location (node) represents the locational value of energy i.e. a cost of supplying extra 1 MW of energy to node. It consists of the cost of energy used at a node and the cost of delivering it there, which depends on losses and congestion. Since each occurrence of congestion leads to graduate increase of delivery cost [5], aggregation (clustering) of similar nodal prices should result in a reliable solution to the problem of division of the market into zones. However, in the literature concerning division into zones [3],[4] usually stable levels of generation are assumed, which remains in contradiction with the increasing amount of renewable generation for which, as yet, wind farms, characterized by highly variable power output, constitute the main source. In our previous work [1] we found that the relative instability in the amount of power injected into the system by wind farms significantly influences the energy prices even if the rate of wind generation to total generation is relatively small. Hence, we extend the clustering of LMP method by taking into account variable weather conditions. In essence, we conduct calculation and clustering of LMP for a range of different wind scenarios, and we add a third, "aggregating" step to the method, namely, the final consensus clustering, which joins single clusterings conducted for each of weather scenarios. The above procedure can be summarized in step-wise fashion as follows:

1. On a sample of historical wind data or Monte Carlo simulations of wind strength, the output of wind farms present in the energy network is estimated.
2. Each scenario of wind output constitutes an input into Direct Current Optimal Power Flow (OPF DC) [6] algorithm, to determine feasible generation in the network and obtain set of nodal prices. In case when congestion arises, the nodal prices will vary for this scenario.[1]
3. For each scenario, nodal prices are clustered using a hierarchical method based on Ward's minimal variance criterion [7], modified to acknowledge the existence of connection (branch) between nodes of the clusters.
4. Since the assignment of nodes into zones can vary from one scenario to another, to derive an "aggregated" division a *consensus clustering* method is used. Specifically, we Cluster-based Similarity Partitioning Algorithm [8] to obtain a tentative division into $k = 1,...,K$ clusters.

The divisions into $k = 1,...,K$ clusters are then to be evaluated by a welfare-maximizing criterion described in section (iii).

**(ii)     Congestion contribution identification**

The congestion contribution identification is a relatively new approach to zonal division [2]. This method for defining a border line that separates zones is based on the analysis of Power Transfer Distribution Factors. The distribution factors reflect the influence of unit nodal injections on power flow along the transmission lines. Thus, grouping the nodes characterized by similar factors into one zone defines a region of desirably similar sensitivity to congestion. As the necessity for zonal division is the congestion limiting power transfer on a transmission line, we focus on PTDF elements corresponding only to the connections which are likely to be congested.

---

[1] Neglecting the losses (a fundament of DC approximation) allows us to concentrate on the variation in nodal prices which arises only as a result of congestion cost.

The specific reasoning is built on a premise that a convenient division is one that minimizes the chance of intra-zonal congestions. Hence, borders of the zones are expected to cross the often congested lines. PTDF elements are helpful in deciding which nodal injection increases, and which decreases an observed congestion. The most problematic task is to concern an inconsistency resulting from arbitrary choice of a reference node while calculating PTDF matrix for the system.

An easy way of dealing with broken symmetry of load that affects distribution factors is to use a special type of PTDF matrix [2].[2] Let us assume that $\mathbf{H}^i \in \mathbb{R}^{M \times N}$ is a PTDF matrix of $M$-line/$N$-bus system built under assumption that $i$ is the reference node. Decision which node we are expected to use as the reference one is crucial when the analysis of separate matrices' elements is concerned, but meaningless as long as we use $\mathbf{H}^i$ matrix only for calculating power flows. In fact all $\mathbf{H}^i$, $i \in \{1,...,N\}$ constitute an equivalence class with respect to left-handed multiplication by vectors $\mathbf{p} = (p_1,..., p_N)^T$ such that

$$\sum_{n=1}^{N} p_n = 0. \tag{1}$$

In other words, for

$$\tilde{\mathbf{p}} = \mathbf{H}^1 \mathbf{p} = \mathbf{H}^2 \mathbf{p} = \ldots = \mathbf{H}^N \mathbf{p}, \tag{2}$$

the product $\tilde{\mathbf{p}}$ remains unaffected if we apply any of $\mathbf{H}^i$ operators to a particular vector $\mathbf{p}$ of nodal injections/withdraws which obviously satisfies the property (1).

Let us prove that adding the same element $\alpha_l$ to selected row of matrix $\mathbf{H}^i$ does not influence the left-hand side:

$$\tilde{p}_l = \sum_{n=1}^{N}(H_{ln}^i + \alpha_l) p_n = \sum_{n=1}^{N} H_{ln}^i p_n + \alpha_l \sum_{n=1}^{N} p_n = \sum_{n=1}^{N} H_{ln}^i p_n. \tag{3}$$

In the consequence, we may choose the coordinates of vector $\boldsymbol{\alpha}$ and create new PTDF operator $\mathbf{S} = \mathbf{H}^i + \boldsymbol{\alpha} \times \mathbf{u}^T$, as the sum of PTDF matrix and a dyadic product of $\boldsymbol{\alpha}$ and $N$-dimensional vector $\mathbf{u}^T = [1, 1, ..., 1]$. In order to choose the proper values $\alpha_l$, we shall discuss the role of distribution factors' signs in the context of choosing a particular reference node.

---

[2] In [2] construction of this special-form PTDF matrix was given without a formal proof, which we present here.

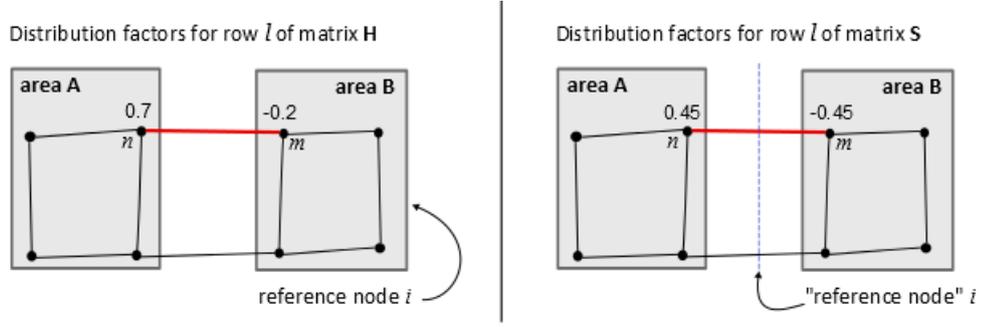

**Figure 1 – Comparison of distribution factors. Their values reflect the position of reference node.**

Let us consider an electric grid consisting of two interconnected areas [Fig 1]. The congested line $l = (n, m)$ defines a row of matrix $\mathbf{H}^i$, elements of which are picked as objects of the following analysis. If distribution factors assigned to nodes $n$ and $m$ are 0.7 and -0.2, respectively, we assume that the reference node $i$ belongs to area B. Changing zonal ascription of $i$ results in the need for readjusting all distribution factors. Meanwhile, there is no consistent method for selecting the right reference node, as in most cases the topology and physical parameters of the grid do not give an opportunity to pick a node that reflects an equivalent influence of power injections at nodes $n$ and $m$ to power flow on the congested line.

In fact, we do not need to decide on a particular node. The solution is to shift all the elements of the row $l$ ($H_{lk}$, $k \in \{1, \ldots, N\}$) by the negative average of the factors reflecting injections in nodes $n$ and $m$. It means that we should consider analyzing the $l$-th row of new operator $\mathbf{S}$

$$\forall k \in \{1,\ldots,N\}, l=(n,m): \quad S_{lk} = H^i_{lk} - \frac{1}{2}\left(H_{lm} + H_{ln}\right). \tag{4}$$

Elements of $\mathbf{S}$ are no longer denoted by $i$, as the differential does not depend of the choice of reference bus for $\mathbf{H}^i$. New distribution factors reflect an imaginary choice of the reference node, which can be treated as situated at electrically equivalent distance form nodes $n$ and $m$. That leads to desired equality of absolute values of PTDFs assigned to both ends of the congested line. Dealing with multiple congestions, the analysis of more than one row of $\mathbf{S}$ is a necessity. In order to give the equation (4) a matrix form, we need to introduce some notation. If $|\mathbf{A}|$ is a matrix of absolute values of line-node incidence matrix denoting a system topology and $\mathrm{diagv}(\cdot)$ is an operator forming a vector from a diagonal line of a given matrix, defined by $[\mathrm{diagv}(\mathbf{M})]_k = M_{kk}$, the generic form of PTDF derived from any regular operator $\mathbf{H}^i$, may be expressed by:

$$\mathbf{S} = \mathbf{H}^i - \frac{1}{2}\mathrm{diagv}\left(\mathbf{H}^i |\mathbf{A}|^T\right) \times \mathbf{u}^T. \tag{5}$$

Then, the nodes can be easily categorized into two zones with respect to a congested line $l$ by simply checking the sign of coefficients $S_{lk}$ for $k \in \{1, ..., N\}$. When taking next congested line, say $l'$, to conduct the division, we can restrain the problem only to a zone which contains $l'$ as an intra-zone link.

The procedure of division into zones using congestion contribution identification can be summarized in step-wise fashion as follows:

1. The generalized PTDF matrix is computed for the network model.
2. On a sample of historical wind data or Monte Carlo simulations of wind strength, the output of wind farms present in the energy network is estimated.
3. Each scenario of wind output constitutes an input into OPF DC algorithm, to determine the lines likely to be congested, that is, for which the transmission limits were a binding conditions in the optimization process.
4. Starting with the most frequently congested line, the network is divided with respect to nodes' congestion contributions, and each of the division is tested according to the welfare criterion to evaluate its necessity. If the division brings betterment, it is applied and further congestions are treated as intra-zonal. If not, this line congestion is left for the balancing market to overcome.

In essence, we provide a simple proof of property referring to matrices S and extend the division methodology described in [2] to take into account lines congested in different wind generation scenarios and to acknowledge the welfare effects of the division.

**(iii)    Welfare criterion**

As the welfare criterion we use the minimization of the overall cost of supplying energy. That is, a specific division of a single market into $k$ zones is considered as welfare-enhancing if the aggregated cost of supplying energy is lower after the division.

The cost of supplying energy on a single market consists of:
- value of the energy traded on the uniform market, derived from the supply-demand equilibrium
- cost of readjustments conducted on balancing market (in order to correct for congestion and energy losses)

The cost of supplying energy on a zonal market consists of:
- value of energy sold at each of $k$ zonal markets, derived from the supply-demand equilibrium taking into account inter-zonal trading (defined by Market Coupling algorithm [9])
- congestion rent on trades between zones
- cost of adjustments conducted on $k$ balancing market (in order to correct for congestion and energy losses)

In order to estimate the costs of balancing market, we first run the OPF DC algorithm on the network model with transmission limits set to infinity, in order to find an unconstrained market equilibrium, with uniform price equal to the highest marginal cost of running generator. Next, we run Alternating Current Optimal Power Flow (OPF AC) with transmission limits, and we compute the costs of adjusting generation with respect to unconstrained OPF DC solution.

The inter-zonal trading ("market coupling") and intra-zone balancing markets are simulated by:

1. Running OPF DC algorithm with constraints only on inter-zonal connections. Zonal prices are determined as highest marginal cost of running generator in a zone.
2. Running separately OPF AC with constraints for sub-network representing each zone. The inflows/outflows on the zone border nodes are adjusted to incorporate inter-zone flows derived by OPF DC above. The costs of adjusting generation with respect to unconstrained OPF DC solution is taken as balancing market charges.

Lastly, the division with highest welfare/lowest cost is treated as the most desirable (recommended) division into zones.